\documentclass[preprint,12pt]{elsarticle}

\usepackage[T1]{fontenc}
\usepackage[utf8]{inputenc}
\usepackage{lmodern}
\usepackage{amsmath,amssymb,amsthm}
\usepackage{bm}
\usepackage{graphicx}
\usepackage{booktabs}
\usepackage{enumitem}
\usepackage{xcolor}
\usepackage[hidelinks]{hyperref}
\usepackage{microtype}
\usepackage{lineno}

\biboptions{sort&compress}
\journal{Computer Methods in Applied Mechanics and Engineering}

\newcommand{\That}{\hat T}
\newcommand{\what}{\hat w}
\newcommand{\Einf}{E_\infty}
\newcommand{\bvarepsilon}{\bm{\varepsilon}}
\newcommand{\bsigma}{\bm{\sigma}}
\newcommand{\btheta}{\bm{\theta}}
\newcommand{\bB}{\mathbf{B}}
\newcommand{\bF}{\mathbf{F}}
\newcommand{\bJ}{\mathbf{J}}
\newcommand{\bW}{\mathbf{W}}
\newcommand{\bu}{\mathbf{u}}
\newcommand{\bff}{\mathbf{f}}
\newcommand{\bv}{\mathbf{v}}
\newcommand{\Pset}{\mathcal{P}}
\newcommand{\Lset}{\mathcal{L}}
\newcommand{\lag}{\mathrm{lag}}
\newcommand{\inst}{\mathrm{inst}}
\newcommand{\dop}{\mathrm{do}}
\newcommand{\pa}{\mathrm{pa}}
\newcommand{\eps}{\varepsilon}

\theoremstyle{definition}
\newtheorem{remark}{Remark}
\theoremstyle{plain}
\newtheorem{proposition}[remark]{Proposition}   

\begin{document}

\begin{frontmatter}

\title{Physics-constrained differentiable causal discovery of constitutive mechanisms through boundary value problems}

\author[ismd]{Michael Kraus\corref{cor1}}
\ead{XXX} 
\cortext[cor1]{Corresponding author.}
\affiliation[ismd]{organization={Institute of Structural Mechanics and Design (ISM+D), Technical University of Darmstadt},
            city={Darmstadt},
            country={Germany}}

\begin{abstract}
Data-driven discovery of constitutive laws selects model terms that reproduce measured responses. When environmental variables such as temperature and moisture co-vary in the data, a well-fitting model can still attribute an effect to the wrong variable and fail under new conditions. We recast constitutive model discovery as causal discovery of mechanism parent sets. A material with internal variables is represented as a time-unrolled structural causal model. Its balance laws and constitutive skeleton are hard-coded; its environment--mechanism and state--state dependencies are learnable edges. Candidate graphs are scored through the boundary value problem by a weak-form equilibrium gap, evaluated by strain-driven integration along full-field displacements, so neither stresses nor internal variables need to be measured. An adaptive lasso selects the edges, and a NOTEARS-type constraint keeps the instantaneous graph acyclic, so that the material update has an explicit order of evaluation. Climate-chamber protocols are treated as interventions. The Gauss--Newton matrix of the score, obtained by automatic differentiation, serves as a local identifiability certificate and drives a greedy D-optimal test design. In a synthetic study on a perforated plate of a hygro-thermo-viscoelastic damaging material (8 true edges among 13 candidates), interventional data recover the true graph exactly in 7 of 10 runs at a displacement noise of about 0.01\,px and make no attribution error. Observational data with a temperature--moisture correlation of $-0.998$ never recover it. Along the confounded directions, their score is 250--380 times flatter. Observationally discovered models reproduce a climate within their training range to 3.6\,\% error but mispredict a hot and humid climate by 16.9\,\% on average and up to 41.5\,\%; interventionally discovered models stay within 3.2\,\%. A correlation sweep shows that the certificate predicts attribution errors but not the noise-induced bias of the weak-form score, which remains the main limitation. The D-optimal design, computed without knowledge of the truth, selects test sets ranked second or third of 126 candidates.
\end{abstract}

\begin{highlights}
\item Constitutive models with internal variables are cast as time-unrolled structural causal models.
\item Mechanism graphs are scored through the boundary value problem without stress data.
\item Climate-chamber protocols act as interventions with a computable identifiability certificate.
\item Observational data misattribute temperature and moisture effects and fail out of distribution.
\item Fisher-based design selects near-optimal tests without knowledge of the truth.
\end{highlights}


\end{frontmatter}

\section{Introduction}\label{sec:intro}

Data-driven discovery of constitutive laws has matured from fitting stress--strain curves to identifying interpretable models directly from full-field displacement data and boundary forces. Unsupervised approaches of the EUCLID family~\cite{Flaschel2021,Flaschel2022,Flaschel2023,Joshi2022} select a sparse model from a library by enforcing equilibrium in weak form, so stress data are never needed. Constitutive artificial neural networks~\cite{LinkaKuhl2023,Linka2021} and sparse regression in the spirit of SINDy~\cite{Brunton2016} follow the same idea: the structure of the law is selected, not prescribed.

These methods answer an associational question. Which candidate terms, active together, reproduce the measured response? For materials whose behaviour depends on environmental and internal variables---temperature $T$, moisture $w$, damage $D$, viscous strain---this is not the question an engineer needs answered. The engineer needs to know which variable changes which mechanism. Does moisture lower the stiffness, or does temperature? Does damage accelerate relaxation, or does relaxation delay damage? Such questions are causal: they ask what happens under an intervention that was not part of the data~\cite{Pearl2009,Peters2017}.

The distinction matters in practice. Laboratory and field data for polymers, timber, concrete or laminated glass interlayers are routinely collected under climates in which $T$ and $w$ co-vary. A model fitted to such data can reproduce every measured curve and still attribute a stiffness loss to the wrong variable. It then fails exactly when it is used for design: under a combination of conditions that was never observed together, for instance hot \emph{and} humid. Low training error is no evidence of correct mechanism attribution. In causal-inference terms, the observational distribution does not identify the interventional one~\cite{Pearl2009,Schoelkopf2021}.

Causal discovery offers the missing vocabulary and the missing guarantees. Structural causal models (SCMs) represent a system as a directed graph of mechanisms~\cite{Pearl2009,Peters2017}. Differentiable formulations such as NOTEARS~\cite{Zheng2018}, DAGMA~\cite{Bello2022} and DYNOTEARS~\cite{Pamfil2020} turn graph search into continuous optimisation. Interventional data shrink the set of graphs compatible with the data, the interventional Markov equivalence class~\cite{Hauser2012,Eberhardt2005}. These methods, however, assume that every node is observed or that latent variables enter through simple noise models. In solid mechanics the decisive variables are fields---stress, damage, viscous strain---that are never measured. They are only accessible through a boundary value problem (BVP).

This paper closes that gap. We treat a constitutive model as a hybrid SCM whose structural equations are partly fixed by balance laws and partly learned, and we score candidate graphs \emph{through} the BVP with a differentiable finite element discretisation. The approach rests on four contributions:
\begin{enumerate}[leftmargin=*]
\item \textbf{Constitutive models as time-unrolled hybrid SCMs.} Balance laws and the constitutive skeleton are hard-coded edges; environment--mechanism and state--state dependencies are learnable edges. Feedback between internal variables is represented across time steps. Acyclicity of the instantaneous subgraph guarantees an explicit order of evaluation of the material update.
\item \textbf{A differentiable score through the BVP.} Candidate graphs are scored by the weak-form equilibrium gap and the reaction-force balance, evaluated by strain-driven integration along measured displacement fields. Only full-field displacements, reaction forces and environmental fields are required.
\item \textbf{Identifiability via interventions.} Climate-chamber protocols that set $T$ and $w$ independently are interpreted as interventions. The Gauss--Newton matrix of the score, computed by automatic differentiation, serves as a local identifiability certificate. It is nearly singular along the collinear directions of observational protocols and well conditioned for interventional ones. A correlation sweep shows what it captures (attribution errors) and what it misses (noise-induced bias).
\item \textbf{Intervention design.} The same matrix drives a greedy D-optimal selection of tests from a candidate library, which turns causal identifiability into an experimental design problem.
\end{enumerate}

A synthetic study on a perforated plate made of a hygro\-/thermo\-/visco\-elastic damaging material demonstrates the approach. The study quantifies structure recovery against noise, the gap between observational and interventional models under a held-out climate, the role of the acyclicity constraint, and the gain from designed interventions.

The paper is organised as follows. Section~\ref{sec:related} reviews related work. Section~\ref{sec:scm} develops the SCM view of constitutive modelling and its identifiability. Section~\ref{sec:method} presents the differentiable discovery algorithm. Section~\ref{sec:experiments} reports the numerical experiments, Section~\ref{sec:discussion} discusses limitations, and Section~\ref{sec:conclusions} concludes.

\section{Related work}\label{sec:related}

\paragraph{Constitutive model discovery without stress data}
Full-field identification has a long history in experimental mechanics, from finite element model updating to the virtual fields method~\cite{Avril2008,Pierron2012}. EUCLID~\cite{Flaschel2021,Flaschel2022,Flaschel2023} combined weak-form equilibrium with sparse regression over a model library. It was extended to plasticity~\cite{Flaschel2022}, generalized standard materials~\cite{Flaschel2023}, linear viscoelasticity~\cite{Marino2023}, Bayesian uncertainty~\cite{Joshi2022} and neural ansatz spaces~\cite{Thakolkaran2022}. Constitutive artificial neural networks~\cite{LinkaKuhl2023,Linka2021} embed thermodynamic and invariance constraints in the network architecture. A recent review is given in~\cite{Fuhg2024}. Common to these works is a fixed set of input variables per mechanism. Environmental dependencies, when present, enter as additional regressors, so the discovered model is a correlational summary of the training protocol.

\paragraph{Differentiable mechanics}
Automatic differentiation frameworks~\cite{Bradbury2018} have made finite element solvers differentiable end to end~\cite{Xue2023}. Implicit differentiation of nonlinear solves~\cite{Blondel2022} and differentiable physics for control and inverse design~\cite{Holl2020} are now routine. We use differentiability in two places: to compute gradients of the discovery score with respect to all mechanism parameters and edge weights, and to compute the Gauss--Newton matrix that certifies local identifiability and drives experimental design.

\paragraph{Causal discovery}
Constraint-based and score-based methods recover a causal graph up to its Markov equivalence class~\cite{Spirtes2000}. NOTEARS~\cite{Zheng2018} expressed acyclicity as a smooth equality constraint, $h(\bW)=\operatorname{tr}\big(\exp(\bW\circ\bW)\big)-d=0$. DAGMA~\cite{Bello2022} replaced it with a log-determinant characterisation, and DYNOTEARS~\cite{Pamfil2020} extended the idea to time series with lagged and contemporaneous edges. Interventional data reduce the equivalence class~\cite{Hauser2012,Eberhardt2005}; DCDI~\cite{Brouillard2020} learns from such data by gradient descent. Benchmarks with standardisation artefacts can make discovery look easier than it is~\cite{Reisach2021}, which is one reason to test against known mechanistic ground truth. The formal link between differential equations and SCMs was established in~\cite{Mooij2013}; cyclic SCMs and their equilibria are treated in~\cite{Bongers2021}. Causal ordering in equilibrium systems goes back to Simon~\cite{Iwasaki1994}.

\paragraph{Causality and physics}
Sun et al.~\cite{Sun2022} discovered causal relations among micro-structural descriptors of granular materials from material-point data generated by discrete element simulations. Chen et al.~\cite{Chen2026} used known ODE dynamics as an inductive bias and learned only the residual causal couplings. Huang et al.~\cite{Huang2023} learned a differentiable causal model of the sim-to-real gap in robotics. In structural health monitoring, environmental variables are recognised as confounders of damage-sensitive features, and their influence is removed by regression~\cite{Neumann2026}.

\paragraph{Gap}
To the best of our knowledge, no method discovers causal structure among constitutive mechanisms when the internal variables are latent fields observed only through a BVP. The same holds for treating environmental test protocols as interventions with a computable identifiability certificate. This paper addresses both.

\section{Constitutive modelling as a structural causal model}\label{sec:scm}

The central idea is that a constitutive model with internal variables is a structural causal model. Its mechanisms are partly known from physics and partly unknown. This section makes the correspondence precise and derives what can and cannot be identified from a given test protocol.

\subsection{Structural causal models}\label{sec:scm-basics}

An SCM consists of endogenous variables $V$, exogenous variables $U$ and one structural assignment per endogenous variable~\cite{Pearl2009,Peters2017}:
\begin{equation}\label{eq:scm}
V_i := f_i\big(\pa(V_i),\, U_i\big), \qquad i = 1,\dots,d .
\end{equation}
The parents $\pa(V_i)$ define a directed graph $G$. An intervention $\dop(V_k = v)$ replaces the assignment of $V_k$ by the constant $v$ and leaves all other mechanisms unchanged. The resulting interventional distribution generally differs from the conditional distribution given $V_k = v$. Two graphs that induce the same observational distribution are Markov equivalent; interventions on selected nodes can break the equivalence~\cite{Hauser2012,Eberhardt2005}.

\subsection{The hybrid mechanism graph}\label{sec:graph}

We consider small-strain, plane-stress kinematics with a scalar-damage, hygro-thermo-viscoelastic material of Zener type~\cite{SimoHughes1998,SimoJu1987,Lemaitre1992}. The variables at a material point and time $t_{n+1}$ are grouped as follows:
\begin{itemize}[leftmargin=*]
\item environment $\mathbf z = (\That, \what)$: normalised temperature $\That = (T - T_0)/20\,\mathrm{K}$ and moisture $\what = (w - w_0)/\Delta w_{\mathrm{ref}}$;
\item kinematics: total strain $\bvarepsilon$, obtained from displacements;
\item mechanism parameters $\mathbf m = (\Einf, E_1, \tau, \kappa_0)$ and eigenstrain $\bvarepsilon_\theta$;
\item latent internal states $\mathbf q = (\bvarepsilon_v, \kappa, D)$: viscous strain, damage history variable and damage;
\item stress $\bsigma$, which is never observed.
\end{itemize}
The structural assignments of one time step are, in their order of evaluation:
\begin{subequations}\label{eq:assignments}
\begin{align}
m_k &:= m_k^0 \exp\!\big(\theta_{kT}\,\That + \theta_{kw}\,\what\big), \qquad k \in \{\Einf, E_1, \kappa_0\},\\
\bvarepsilon_\theta &:= 10^{-3}\big(\theta_{\theta T}\,\That + \theta_{\theta w}\,\what\big)\,(1,1,0)^{\mathsf T}, \qquad \bvarepsilon_m := \bvarepsilon - \bvarepsilon_\theta,\\
\tau &:= \tau^0 \exp\!\big(\theta_{\tau T}\,\That + \theta_{\tau w}\,\what + \theta^{\lag}_{\tau D}\, D^{n} + \theta^{\inst}_{\tau D}\, D^{n+1}\big),\label{eq:tau}\\
\bvarepsilon_v^{n+1} &:= \frac{\bvarepsilon_v^{n} + (\Delta t/\tau)\,\bvarepsilon_m}{1+\Delta t/\tau},\\
\bvarepsilon_d &:= \bvarepsilon_m - \theta_{Dv}\,\bvarepsilon_v^{n+1}, \qquad \kappa^{n+1} := \max\big(\kappa^{n}, \lVert\bvarepsilon_d\rVert\big),\\
D^{n+1} &:= D_{\max}\Big(1 - \exp\big(-\langle \kappa^{n+1}-\kappa_0\rangle / \kappa_f\big)\Big),\\
\bsigma &:= (1-D^{n+1})\,\bar{\mathbf C}\big[\Einf\,\bvarepsilon_m + E_1(\bvarepsilon_m - \bvarepsilon_v^{n+1})\big].
\end{align}
\end{subequations}
Here $\bar{\mathbf C}$ is the plane-stress elasticity matrix for unit modulus and Poisson ratio $\nu$, $\langle\cdot\rangle$ the Macaulay bracket, and $\lVert\bvarepsilon_d\rVert^2 = \eps_{xx}^2 + \eps_{yy}^2 + \gamma_{xy}^2/2$. The viscous update is a backward-Euler step of $\dot{\bvarepsilon}_v = (\bvarepsilon_m - \bvarepsilon_v)/\tau$.

Two kinds of edges are distinguished. \textbf{Hard edges} encode physics that is not in question: kinematics $\bvarepsilon = \bB\bu$, balance of linear momentum in weak form, and the constitutive skeleton for $\bsigma$. \textbf{Learnable edges} are the 13 coefficients $\theta$ listed in Table~\ref{tab:edges}. An edge $j \to k$ is present if and only if $\theta_{kj} \neq 0$. The discovery problem is to find the support of $\btheta$, which is the causal graph, together with its values and the base parameters $\mathbf m^0$, $\tau^0$, $\kappa_f$, $D_{\max}$.

\begin{table}[t]
\centering
\caption{Candidate learnable edges and their ground-truth values in the synthetic study (Section~\ref{sec:experiments}). Physical edges such as environment $\to$ environment or mechanism $\to$ environment are excluded by construction, because $T$ and $w$ are exogenous in a controlled test.}
\label{tab:edges}
\small
\begin{tabular}{@{}rllr@{}}
\toprule
\# & Edge & Mechanism & Ground truth $\theta$ \\
\midrule
0 & $T \to \Einf$ & thermal softening of equilibrium modulus & $-0.5$ \\
1 & $w \to \Einf$ & moisture softening of equilibrium modulus & $-0.3$ \\
2 & $T \to E_1$ & thermal softening of non-equilibrium modulus & $0$ \\
3 & $w \to E_1$ & moisture softening of non-equilibrium modulus & $0$ \\
4 & $T \to \tau$ & thermally activated relaxation & $-1.0$ \\
5 & $w \to \tau$ & moisture-activated relaxation & $0$ \\
6 & $T \to \kappa_0$ & thermal effect on damage threshold & $0$ \\
7 & $w \to \kappa_0$ & moisture-reduced damage threshold & $-0.4$ \\
8 & $T \to \bvarepsilon_\theta$ & thermal expansion ($\times 10^{-3}$) & $1.4$ \\
9 & $w \to \bvarepsilon_\theta$ & hygric swelling ($\times 10^{-3}$) & $2.0$ \\
10 & $\bvarepsilon_v \to D$ & viscous strain reduces damage driving strain & $0.6$ \\
11 & $D^n \to \tau$ & damage-accelerated relaxation, lagged & $-2.0$ \\
12 & $D^{n+1} \to \tau$ & damage-accelerated relaxation, instantaneous & $0$ \\
\bottomrule
\end{tabular}
\end{table}

\subsection{Time-unrolled graph and acyclicity}\label{sec:acyclicity}

Mechanics is full of feedback: damage reduces stress, stress drives damage, and damage may accelerate relaxation, which in turn changes the strain that drives damage. A static DAG cannot represent such loops. Following DYNOTEARS~\cite{Pamfil2020}, we split the weighted adjacency into an instantaneous part $\bW_0$ (edges within one time step) and a lagged part $\bW_1$ (edges from $t_n$ to $t_{n+1}$). History variables ($\kappa$, $\bvarepsilon_v$, $D$ at $t_n$) are the lagged parents. Feedback is then admissible as long as every loop passes through at least one lag.

\begin{remark}[Evaluation order]\label{rem:order}
Conditional on the measured strain, the assignments of one time step can be evaluated in a single pass, in a fixed order, if the instantaneous graph $\bW_0$ restricted to the latent states is acyclic, because a topological order then exists. If $\bW_0$ contains a cycle, the assignments form a simultaneous system, i.e.\ a cyclic SCM~\cite{Bongers2021}, which in general requires a local fixed-point or Newton iteration. Some simultaneous systems have closed-form solutions: the backward-Euler viscous update is itself an implicit equation in $\bvarepsilon_v$ that we solve in closed form. Acyclicity is therefore sufficient for a single-pass update but not necessary for solvability. We use it as a modelling choice. It keeps the order of evaluation explicit and places all feedback between latent states across time.
\end{remark}

In the candidate set of Table~\ref{tab:edges}, the instantaneous subgraph over $(\bvarepsilon_v, D)$ contains edge 10 ($\bvarepsilon_v \to D$) and edge 12 ($D^{n+1} \to \tau \to \bvarepsilon_v$). We enforce acyclicity with the NOTEARS function~\cite{Zheng2018}:
\begin{equation}\label{eq:notears}
h(\bW_0) = \operatorname{tr}\!\big(\exp(\bW_0 \circ \bW_0)\big) - d, \qquad \bW_0 = \begin{pmatrix} 0 & s_{10}\,\theta_{Dv} \\ s_{12}\,\theta^{\inst}_{\tau D} & 0 \end{pmatrix},
\end{equation}
with $d = 2$ and effect scales $s_k$ (Section~\ref{sec:sparsity}). Then $h = 2\cosh\big(s_{10}\theta_{Dv}\, s_{12}\theta^{\inst}_{\tau D}\big) - 2$,\linebreak[2] which vanishes if and only if one of the two edges is absent. Remark~\ref{rem:order} gives the constraint a mechanical reading. It selects models whose material update has an explicit order of evaluation, and it places any feedback across time, where history variables live. This is consistent with Simon's causal ordering of dynamical systems~\cite{Mooij2013,Iwasaki1994}.

Two levels of the model must be distinguished. The SCM of Section~\ref{sec:graph} is a material-point model \emph{conditional on the strain}. At the level of the boundary value problem, strain is itself an effect: $D \to \bsigma \to \text{(equilibrium)} \to \bu \to \bvarepsilon \to D$ is an instantaneous global loop, an equilibrium system in the sense of Simon~\cite{Iwasaki1994} and of cyclic SCMs~\cite{Bongers2021}. An intervention on $T$ therefore changes $\bvarepsilon$ through the BVP. The discovery score of Section~\ref{sec:method} treats the measured strain as an observed parent and evaluates each mechanism conditional on it. This cuts the global loop in the same way that score-based causal discovery regresses each node on its observed parents. Interventional predictions (Section~\ref{sec:prediction}) must instead solve the full cyclic system, which is why they need the Newton solver.

\subsection{Test protocols as interventions}\label{sec:interventions}

In a material test, the environment is either \emph{observed} or \emph{set}. In outdoor or uncontrolled laboratory conditions, a common driver $U$ (weather, season, conditioning history) causes both $T$ and $w$. The local fields then obey $w \approx cT + b$ with $c < 0$: warm air is dry. In a climate chamber, the experimenter sets $T$ and $w$ independently. In SCM terms, this is the intervention $\dop\big(T = T^*(\mathbf x,t),\ w = w^*(\mathbf x,t)\big)$, which removes the edges $U \to T$ and $U \to w$.

The observational regime does not confound the mechanism itself; it violates \emph{positivity} (overlap). For the effect of $T$ on $\Einf$, moisture lies on the back-door path $T \leftarrow U \to w \to \Einf$ and must be adjusted for. Adjustment requires variation of $T$ at fixed $w$, which the observational climate does not provide~\cite{Pearl2009,Peters2017}. A designed protocol restores positivity by construction, exactly as randomisation does in a clinical trial.

Because $w$ is measured, the difficulty of the observational regime is lack of overlap, i.e.\ collinearity of the parents, not a hidden confounder of a mechanism. In regression terms, the lasso selects the wrong parent when the irrepresentable condition is violated~\cite{ZhaoYu2006,Meinshausen2006}. The variance inflation factor $1/(1-\rho^2)$ quantifies the loss of information; it is about 209 for the correlation $\rho = -0.998$ of Section~\ref{sec:experiments}. We nevertheless keep the causal vocabulary, for three reasons. The question is parent-set identification for use under intervention. Test protocols are interventions with known targets. And the timing of state--state edges (Section~\ref{sec:acyclicity}) is a genuine question of graph structure, not of regressor choice. The causal setting is deliberately the simplest one, with all edge orientations from the environment fixed by physics; Section~\ref{sec:discussion} discusses hidden confounders.

\subsection{Local identifiability}\label{sec:identifiability}

Let $\mathbf r(\btheta)$ be the stacked, normalised residual vector of the discovery score (Section~\ref{sec:score}) for a protocol $\Pset$. With noise-free data $\mathbf r(\btheta^*) = \mathbf 0$ at the true parameters. Local identifiability at $\btheta^*$ is governed by the Gauss--Newton matrix
\begin{equation}\label{eq:fisher}
\bF_{\Pset}(\btheta^*) = \bJ^{\mathsf T}\bJ, \qquad \bJ = \frac{\partial \mathbf r}{\partial \btheta}\Big|_{\btheta^*}.
\end{equation}
This is the Gauss--Newton matrix of the score, computed exactly by forward-mode automatic differentiation. A zero eigenvalue identifies a direction in parameter space along which the score cannot distinguish models~\cite{Raue2009}. Throughout the paper all residuals of all tests are normalised by one fixed reference force $F_G$, the largest reaction over all simulated tests. Matrices of different tests and protocols are then additive and comparable. $\bF$ is a property of the score and is used as a \emph{local identifiability certificate}. It is not the Fisher information of the measurement model: displacement noise enters the residual nonlinearly, through strains and the history recursion, so $\bJ^{\mathsf T}\bJ$ yields no Cram\'er--Rao bound.

\begin{proposition}[Collinear environment]\label{prop:collinear}
Suppose every test of $\Pset$ satisfies $\what(\mathbf x,t) = c\,\That(\mathbf x,t) + b$ with constants $c$ and $b$ common to all tests and times. For $k \in \{\Einf, E_1, \tau, \kappa_0\}$, let $\bv_k$ change $(\theta_{kT}, \theta_{kw})$ by $(-c, 1)$ and $\log m_k^0$ by $-b$, with zeros elsewhere. Then $\bJ\bv_k = \mathbf 0$. If $b = 0$, the same holds for the eigenstrain direction $\bv_{\eps\theta}$, which changes $(\theta_{\theta T}, \theta_{\theta w})$ by $(-c, 1)$. Hence $\operatorname{rank}\bF_{\Pset} \le \dim\btheta - 4$, and $\le \dim\btheta - 5$ for $b = 0$.
\end{proposition}

\begin{proof}
Along $\bv_k$, the exponent $\theta_{kT}\That + \theta_{kw}\what + \log m_k^0$ changes by $\epsilon(-c\That + \what - b) = 0$. For the eigenstrain, the change is $\epsilon\cdot 10^{-3} b$, which vanishes for $b = 0$. All structural assignments, and hence all residuals, are unchanged.
\end{proof}

If the offset varies between tests or in time, $b = b_p(t)$, no exact null direction remains, and identifiability rests on the variation of $b_p$ alone. In protocol O of Section~\ref{sec:experiments}, $b_p(t) = s(t)\,\delta_p$ with $|\delta_p| \le 0.05$. The directions $\bv_k$ without base components are then close to the flattest directions, and we use them in Fig.~\ref{fig:fisher}b. Proposition~\ref{prop:collinear} is the mechanics analogue of a positivity violation (Section~\ref{sec:interventions}). It explains why a sparse fit to observational data does not fail randomly. It attributes the combined effect to one parent, chosen by noise and regularisation rather than by physics. Section~\ref{sec:recovery} shows that noise adds a second, design-dependent failure mode: bias.

\section{Differentiable causal discovery through the boundary value problem}\label{sec:method}

The algorithm scores a candidate graph by how well it balances forces in the measured specimen, then selects a sparse, acyclic graph in three stages. Every step is differentiable, and no forward solve is needed inside the optimisation.

\subsection{Weak-form score}\label{sec:score}

The specimen is discretised with linear triangles. For each test $p$ and time step $n$ the measured nodal displacements $\hat\bu^n$ give element strains $\bvarepsilon_e^n = \bB_e\hat\bu^n$. Integrating the candidate SCM along these strains (Section~\ref{sec:integration}) yields stresses $\bsigma_e^n(\btheta)$ and the internal force vector
\begin{equation}\label{eq:fint}
\bff^{n}(\btheta) = \sum_{e} A_e\, \bB_e^{\mathsf T}\, \bsigma_e^{n}(\btheta).
\end{equation}
At free degrees of freedom no external load acts, so equilibrium requires $\bff_{\mathrm{free}} = \mathbf 0$. On the loaded edge only the total reaction $\hat F^n$ is measured. The score combines both conditions, as in EUCLID~\cite{Flaschel2021}:
\begin{equation}\label{eq:score}
S(\btheta) = \frac{1}{N_p N_t} \sum_{p=1}^{N_p} \sum_{n=1}^{N_t} \left[ \frac{1}{n_{\mathrm{free}}} \Big\lVert \frac{\bff^{n}_{\mathrm{free}}}{F_{\mathrm{node}}} \Big\rVert^2 + \lambda_F \Big( \frac{\sum_{i \in \Gamma_{\mathrm{top}}} f_i^{n} - \hat F^{n}}{F_{\mathrm{tot}}} \Big)^2 \right].
\end{equation}
$F_{\mathrm{tot}}$ is the largest measured reaction over all tests, $F_{\mathrm{node}} = F_{\mathrm{tot}}/n_{\mathrm{top}}$, and $\lambda_F = 1$. Stresses never enter the score directly. The latent fields $\bvarepsilon_v$, $\kappa$ and $D$ are reconstructed implicitly, through the structural equations.

\subsection{Strain-driven integration}\label{sec:integration}

Given the measured strain history of an element, the material update of Section~\ref{sec:graph} is an explicit recursion in the internal states. We implement it as a \texttt{lax.scan} over time, vectorised over elements and tests with \texttt{vmap} in JAX~\cite{Bradbury2018}. Gradients of $S$ with respect to all 19 parameters (6 base parameters, 13 edge weights) follow from reverse-mode automatic differentiation through the recursion. The cost of one score evaluation is linear in the number of elements, time steps and tests.

If both instantaneous edges 10 and 12 are active, the update contains an algebraic loop (Remark~\ref{rem:order}). We then perform $N_{\mathrm{fp}}$ local fixed-point sweeps per step, starting from $D^n$: $N_{\mathrm{fp}} = 2$ in the main runs and $N_{\mathrm{fp}} = 8$ in the ablation of Section~\ref{sec:ablation}, where the loop stays active. The sweeps are exact whenever edge 12 is absent, which the acyclicity constraint enforces at convergence.

A Newton solver is needed only for two purposes: to generate synthetic data and to predict the response of a discovered model under a new protocol. It assembles element tangents obtained by forward-mode differentiation of the element force and solves the dense linear system for the free degrees of freedom.

\subsection{Sparsity and acyclicity}\label{sec:sparsity}

Edge weights act on parents of different magnitude. To compare them, each edge $k$ carries an effect scale $s_k$, so that $s_k|\theta_k|$ is the typical change of the child's exponent or strain. We use $s_k = 1$ for the environmental edges, whose parents are normalised to about $[-1,1]$ ($|\That|, |\what| \le 1.1$ including the gradients), and for $\bvarepsilon_v \to D$, whose weight $\theta_{Dv}$ is the dimensionless fraction of the viscous strain subtracted from the driving strain. For the damage edges we use $s_k = 0.5$, since $D \le D_{\max} \approx 0.5$.

The selection proceeds in three stages:
\begin{enumerate}[leftmargin=*]
\item \textbf{Full-graph fit.} Minimise $S/S_0$ over all parameters with L-BFGS~\cite{LiuNocedal1989,NocedalWright2006}, starting from base parameters $\btheta_0$ and all edges zero. Here $S_0 = S(\btheta_0)$. The result is $\hat\btheta^{(1)}$ with score $S_1$.
\item \textbf{Adaptive lasso with acyclicity.} Minimise
\begin{equation}\label{eq:stage2}
\begin{split}
&\frac{S(\btheta)}{S_1 + \epsilon S_0} + \lambda \sum_{k} \omega_k \sqrt{(s_k\theta_k)^2 + 10^{-8}} + \mu\, h(\bW_0) + \frac{\rho}{2}\, h(\bW_0)^2,\\
&\omega_k = \frac{1}{s_k|\hat\theta^{(1)}_k| + \delta},
\end{split}
\end{equation}
with the adaptive weights of Zou~\cite{Zou2006}. The augmented-Lagrangian parameters follow $\rho \in \{1, 10, 100\}$ and $\mu \leftarrow \mu + \rho h$ after each inner solve.
\item \textbf{Prune and refit.} Remove edges with $s_k|\theta_k| < \eta$ and refit the remaining parameters without penalty, which removes the shrinkage bias. If an instantaneous 2-cycle survives stage 2, the weaker of its two edges is removed. This mirrors the thresholding step of DAGMA~\cite{Bello2022}.
\end{enumerate}
Normalising the data term by the stage-1 score $S_1$ makes $\lambda$ a dimensionless trade-off. It measures the relative loss increase that a unit of weighted edge magnitude must buy. The floor $\epsilon = 10^{-6}$ keeps the normalisation finite for noise-free data. Unless stated otherwise, $\lambda = 10^{-2}$, $\delta = 0.05$ and $\eta = 0.1$, with 150, $3\times 80$ and 150 L-BFGS iterations in the three stages.

\subsection{Fisher-based intervention design}\label{sec:oed}

For a set of tests $\Pset$, the Gauss--Newton matrix is additive over tests, $\bF_{\Pset} = \sum_p \bF_p$, where each $\bF_p = \bJ_p^{\mathsf T}\bJ_p$ is computed with \texttt{jacfwd} through the score. Given data already collected (protocol $\Pset_0$) and a library $\Lset$ of feasible climate-chamber tests, we select $K$ tests greedily by the D-optimality criterion~\cite{Atkinson2007,HuanMarzouk2013,Huan2024}:
\begin{equation}\label{eq:greedy}
p_{j+1} = \arg\max_{p \in \Lset \setminus \Pset_j} \; \log\det\!\Big(\bF_{\Pset_0} + \sum_{q \in \Pset_j} \bF_q + \bF_p + \epsilon_F \mathbf I\Big).
\end{equation}
All $\bF_p$ use the common reference force of Section~\ref{sec:identifiability}, so that the criterion does not depend on which tests are combined. The matrices are evaluated at the current estimate $\hat\btheta$, obtained from $\Pset_0$ alone, which makes the design sequential and avoids any knowledge of the truth. In practice $\bF_p$ requires full-field data of test $p$. During design, these are replaced by a forward prediction of the current model; Section~\ref{sec:design} discusses this choice.

\subsection{Implementation}

The implementation uses JAX with double precision on a CPU and optax for L-BFGS. The specimen, the discovery score, the Newton solver and the experiments together comprise about 600 lines of Python, available with this paper.

\section{Numerical experiments}\label{sec:experiments}

All experiments use synthetic data from a known ground-truth SCM. This is the only setting in which structure recovery can be measured exactly; Section~\ref{sec:discussion} discusses the step to laboratory data.

\subsection{Setup}\label{sec:setup}

\begin{figure}[t]
\centering
\includegraphics[width=\textwidth]{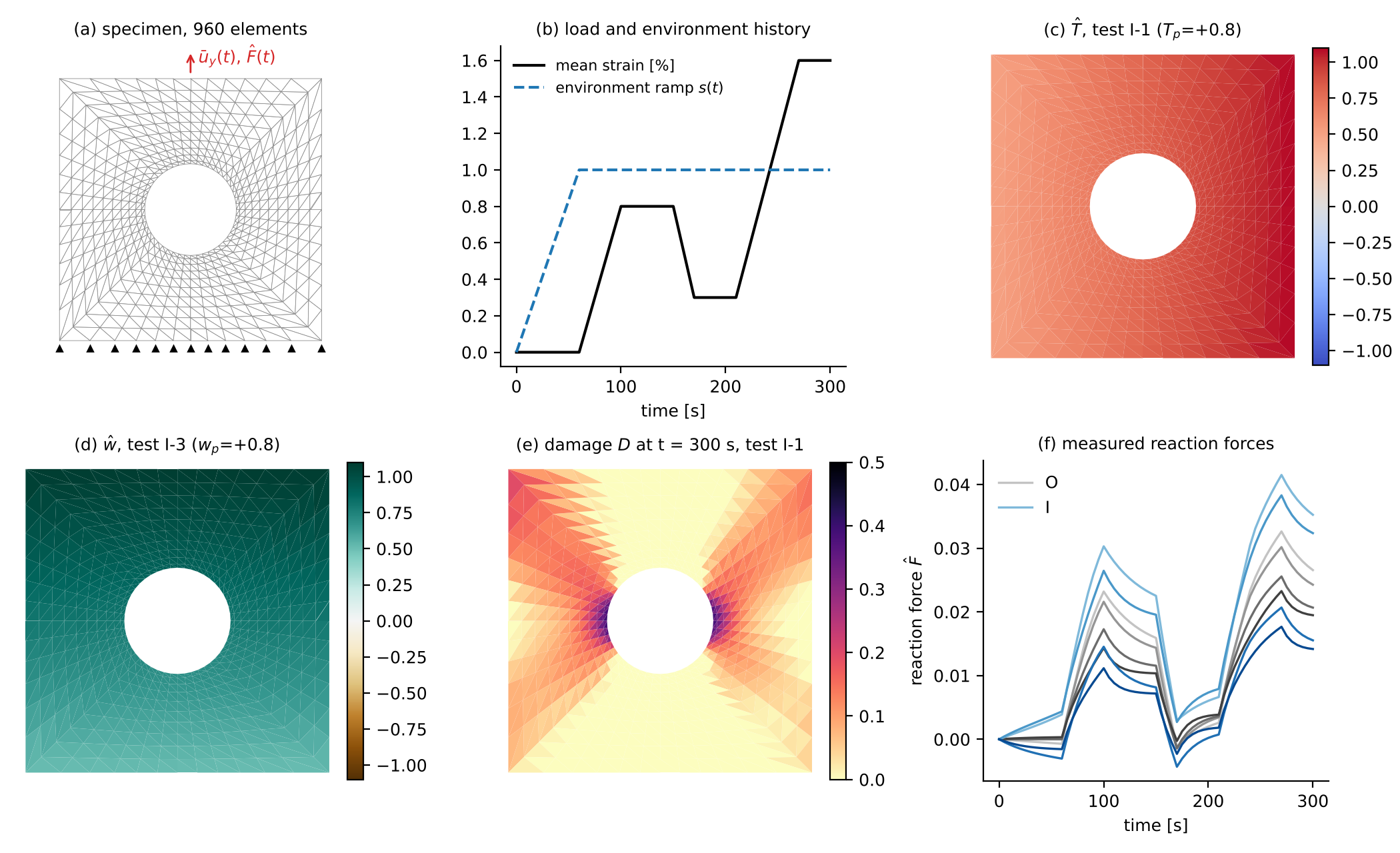}
\caption{Synthetic benchmark. (a) Perforated plate with 960 linear triangles and its boundary conditions. (b) Mean axial strain and environmental ramp $s(t)$. (c) Temperature field of interventional test I-1 ($T_p = +0.8$). (d) Moisture field of test I-3 ($w_p = +0.8$). (e) Damage at $t = 300$\,s in test I-1. (f) Reaction forces of the four observational (grey) and four interventional (blue) tests.}
\label{fig:setup}
\end{figure}

\paragraph{Specimen and discretisation}
A square plate of half-width 1 with a central circular hole of radius 0.35 is modelled in plane stress (Fig.~\ref{fig:setup}). A structured polar mesh, graded towards the hole, has 528 nodes, 960 linear triangles and 1056 degrees of freedom. The bottom edge is fixed in $y$, its midpoint also in $x$. The top edge carries a prescribed vertical displacement and is free in $x$. Only the total vertical reaction at the top edge is recorded as a force.

\paragraph{Load and environment history}
Each test lasts 300\,s with 60 steps of $\Delta t = 5$\,s. During a conditioning phase (0--60\,s) the plate is clamped at zero displacement while the environment ramps linearly to its target. Thermal and hygric eigenstrains therefore build up eigenstresses that carry information on edges 8 and 9. The mean axial strain then follows a ramp to 0.8\,\% (100\,s), a hold (150\,s), an unloading to 0.3\,\% (170\,s), a hold (210\,s), a reloading to 1.6\,\% (270\,s) and a final hold (300\,s). Holds excite relaxation, and the reloading drives damage at the hole.

\paragraph{Ground truth}
The true SCM uses the edges of Table~\ref{tab:edges} with $\Einf^0 = 1$, $E_1^0 = 1.5$ (moduli relative to a reference modulus), $\tau^0 = 20$\,s, $\kappa_0^0 = 0.006$, $\kappa_f = 0.01$, $D_{\max} = 0.5$ and $\nu = 0.35$, which is assumed known. With $\That = 1$ corresponding to $+20$\,K, edge 8 gives a thermal expansion coefficient of $7\times 10^{-5}\,\mathrm{K}^{-1}$, a typical value for polymers. Damage reaches 0.39--0.46 at the hole in all tests.

\paragraph{Protocols}
Environmental fields are prescribed per element centroid $(x,y)$ as $\That = s(t)(T_p + 0.3x)$, where $s$ ramps from 0 to 1 during conditioning.
\begin{itemize}[leftmargin=*]
\item \emph{Observational protocol O} (4 tests, field-like climate): $T_p \in \{-0.8,\allowbreak -0.3,\allowbreak 0.3,\allowbreak 0.8\}$ and $\what = -0.8\,\That + s(t)\,\delta_p$, with small offsets $\delta_p \in \{0.04,\allowbreak -0.05,\allowbreak 0.03,\allowbreak -0.02\}$. The Pearson correlation of $\That$ and $\what$ over all elements and times is $-0.998$.
\item \emph{Interventional protocol I} (4 tests, climate chamber): $(T_p, w_p) \in \{(\pm 0.8, 0),\allowbreak (0, \pm 0.8)\}$ with $\what = s(t)(w_p + 0.3y)$. The correlation is 0.000 by design.
\item \emph{Library $\Lset$} (9 tests): the $3\times 3$ factorial $T_p, w_p \in \{-0.8, 0, 0.8\}$ with the gradients of I.
\item \emph{Held-out tests:} H$_{\mathrm{in}}$ ($T_p = 0.6$ on the observational manifold, i.e.\ hot and dry) and H$_{\mathrm{out}}$ $= (0.8, 0.8)$, hot and humid, a climate the observational protocol never shows.
\end{itemize}

\paragraph{Noise}
Independent Gaussian noise with standard deviation $\sigma_u \in \{0, 10^{-5}, 3\times 10^{-5}\}$ is added to all unconstrained nodal displacements, at every step. For a 100\,mm wide specimen imaged at 2000 pixels, $\sigma_u = 10^{-5}$ corresponds to 0.5\,\textmu m or about 0.01\,px, typical for digital image correlation. The maximal top displacement is 0.032. The reaction force receives noise of 0.5\,\% of its maximum.

\paragraph{Runs and metrics}
Each configuration is repeated with ten seeds at $\sigma_u = 10^{-5}$ and five seeds at $3\times 10^{-5}$; noise-free data are run once. A seed draws the noise and adds $\mathcal N(0, 0.3^2)$ to each base parameter of the initial guess in its own parameterisation (the logarithm, or the logit for $D_{\max}$); all edges start at zero. Structure recovery is measured by the structural Hamming distance (SHD) between the recovered and the true support over the 13 candidate edges. We also report edge inclusion frequencies, parameter errors and prediction errors on the held-out tests.

\subsection{Identifiability}\label{sec:ident-results}

\begin{figure}[t]
\centering
\includegraphics[width=\textwidth]{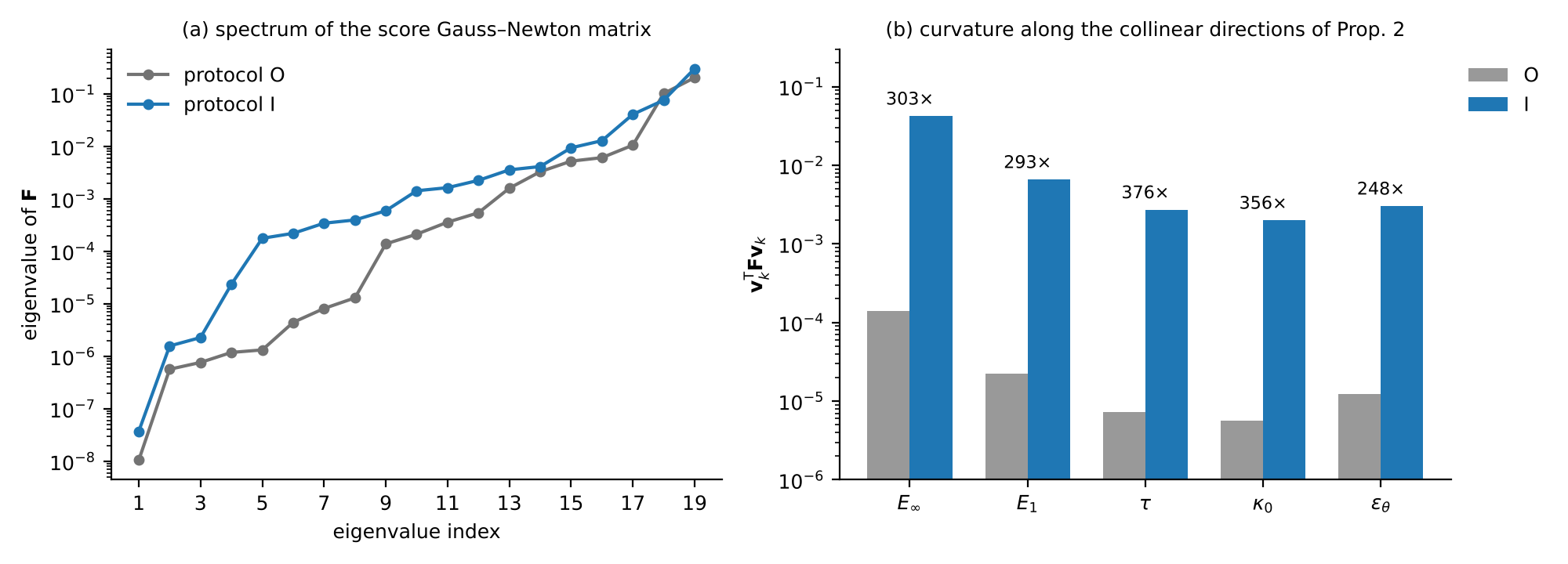}
\caption{Local identifiability at the true parameters, noise-free data, common reference force for all tests. (a) Eigenvalues of the score Gauss--Newton matrix for protocols O and I. (b) Curvature along the collinear directions $\bv_k$ of Proposition~\ref{prop:collinear}; the labels give the ratio between protocols I and O.}
\label{fig:fisher}
\end{figure}

Along the collinear directions of Proposition~\ref{prop:collinear}, the score of the observational protocol is 250 to 380 times flatter than that of the interventional protocol. Both protocols use four tests and the same mechanical history (Fig.~\ref{fig:fisher}). We computed $\bF = \sum_p \bJ_p^{\mathsf T}\bJ_p$ at the true parameters from noise-free data, with the common reference force of Section~\ref{sec:identifiability}.

Fig.~\ref{fig:fisher}a compares the two spectra. Both share one near-zero eigenvalue ($1.0\times 10^{-8}$ for O, $3.7\times 10^{-8}$ for I). Its eigenvector is almost exactly the difference between the lagged and the instantaneous damage edge, with components of magnitude 0.64 and 0.77 for O and 0.68 and 0.74 for I. With $\Delta t = 5$\,s, $D^n$ and $D^{n+1}$ are too similar for any protocol to separate the two edges. This ambiguity is structural and does not depend on the environment. It is resolved by the acyclicity constraint (Section~\ref{sec:ablation}), not by data. Apart from this pair, protocol O has seven eigenvalues below $10^{-4}$, protocol I three.

Fig.~\ref{fig:fisher}b isolates the directions $\bv_k$ of Proposition~\ref{prop:collinear}, with $c = -0.8$, and shows the curvature $\bv_k^{\mathsf T}\bF\bv_k$. Under protocol O these directions are flatter than under protocol I by factors of 303 for $\Einf$, 293 for $E_1$, 376 for $\tau$, 356 for $\kappa_0$ and 248 for $\bvarepsilon_\theta$. These factors are of the order of the variance inflation factor $1/(1-\rho^2) \approx 209$, as expected for collinear parents. In the orthogonal direction $(\theta_{kT}, \theta_{kw}) \propto (1, c)$, which measures the combined effect, protocol O is as informative as protocol I: 0.094 against 0.081 for $\Einf$. The observational data thus determine \emph{how much} the climate changes each mechanism, but hardly \emph{through which variable}. The square roots of the diagonal of $\bF^{-1}$ for the environmental edges are 8 to 18 times larger for O than for I. For the edge $\bvarepsilon_v \to D$, which does not involve the environment, the factor is only 1.5.

\paragraph{Correlation sweep}
To separate the effect of correlation from the specific protocol, we generated three further observational protocols, O$_a$, with larger offsets $\delta_p = a\,(0.8, -1.0, 0.6, -0.4)$ for $a \in \{0.2, 0.6, 1.0\}$. They have $\rho = -0.967$, $-0.822$ and $-0.698$ (protocol O has $a = 0.05$). Within each test, $\what$ remains an affine function of $\That$ in space; decorrelation comes only from the offsets between tests. Along $\bv_{\kappa_0}$, the curvature rises from $5.6\times 10^{-6}$ (O) to $9.3\times 10^{-5}$, $8.4\times 10^{-4}$ and $2.3\times 10^{-3}$. The last value exceeds that of protocol I ($2.0\times 10^{-3}$), and the other mechanisms behave alike. By this certificate, protocol O$_{1.0}$ is as identifiable as protocol I. Section~\ref{sec:recovery} shows that it nevertheless fails under noise, so the certificate is necessary but not sufficient.

\subsection{Structure recovery and noise}\label{sec:recovery}

\begin{figure}[t]
\centering
\includegraphics[width=\textwidth]{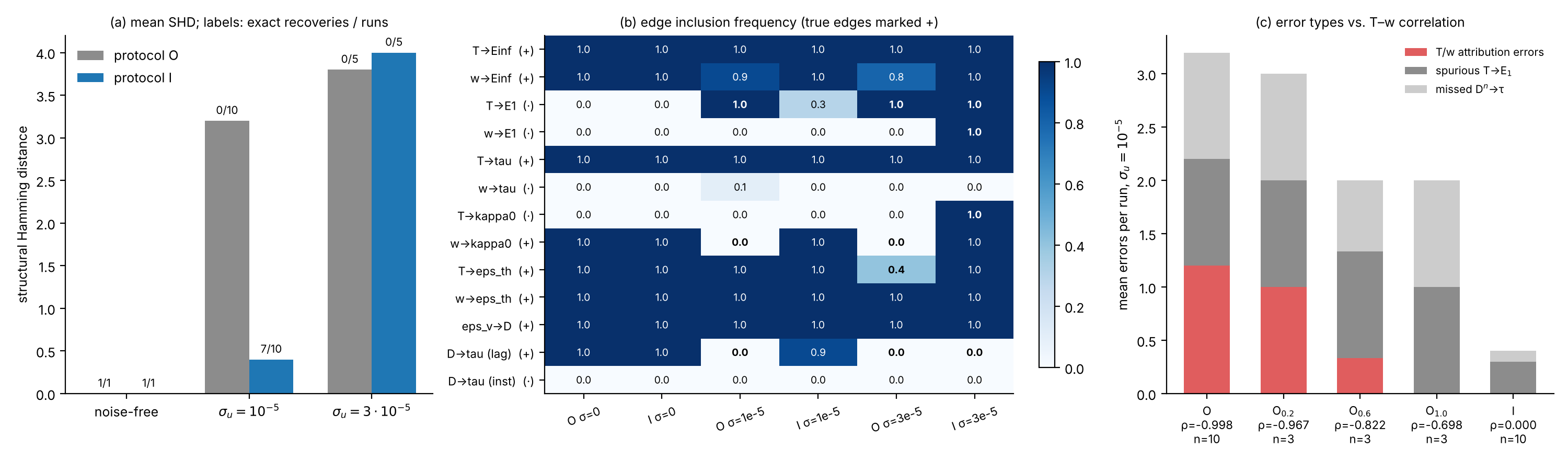}
\caption{Structure recovery. (a) Mean structural Hamming distance per protocol and noise level; labels give exact recoveries per number of runs. (b) Edge inclusion frequency; bold numbers mark cells where the majority decision differs from the ground truth. (c) Mean number of errors per run at $\sigma_u = 10^{-5}$ by type, for protocol O, the correlation sweep O$_a$ (three seeds each) and protocol I.}
\label{fig:recovery}
\end{figure}

At a realistic DIC noise level of $\sigma_u = 10^{-5}$, the interventional protocol recovers the true graph exactly in 7 of 10 runs (mean SHD 0.4). The observational protocol does so in none (mean SHD 3.2). Fisher's exact test gives $p = 0.003$ for this difference. Fig.~\ref{fig:recovery} summarises all runs.

\paragraph{Noise-free data}
Both protocols recover the exact graph (SHD 0). The interventional estimates match the truth to three digits; for example, $\theta = -0.500$, $-0.300$, $-1.000$ and $-0.400$ for edges 0, 1, 4 and 7. Without noise, the small offsets $\delta_p$ suffice to separate $T$ from $w$ even in protocol O, in line with Section~\ref{sec:ident-results}.

\paragraph{$\sigma_u = 10^{-5}$: three types of error}
The errors fall into three types (Fig.~\ref{fig:recovery}c).
\begin{enumerate}[leftmargin=*]
\item \emph{Attribution errors} between $T$ and $w$, as predicted by Proposition~\ref{prop:collinear}. Protocol O misses $w \to \kappa_0$ in all ten runs, misses $w \to \Einf$ in one and adds $w \to \tau$ in one. Protocol I makes no attribution error in any of its ten runs ($p = 1\times 10^{-5}$). The expansion edges illustrate the mechanism. Under O, their estimates scatter with standard deviations of 0.29 and 0.35 over ten runs, against 0.012 and 0.011 under I. Their combination $\theta_{\theta T} + c\,\theta_{\theta w} = -0.19$ nevertheless matches the truth of $-0.20$.
\item \emph{A spurious edge} $T \to E_1$ with $|\theta| \approx 0.15$, in all ten O runs and in three I runs. It compensates a noise-induced bias of the viscous branch: $E_1^0 \approx 1.30$--$1.34$ instead of 1.5 and $\tau^0 \approx 23$\,s instead of 20\,s under both protocols. Its magnitude lies just above the pruning threshold $\eta = 0.1$.
\item \emph{Loss of the lagged damage edge} $D^n \to \tau$, in all ten O runs and one I run. This edge has the smallest curvature of all true edges (Section~\ref{sec:ident-results}). Under O its loss accompanies the misattributed damage threshold. Where the edge is retained under I (9 of 10 runs), its value averages $-5.0$ with a standard deviation of 1.6, against a truth of $-2.0$. Its detection is reliable; its value is not.
\end{enumerate}

\paragraph{Threshold sensitivity}
We recomputed the support from the stage-2 estimates for $\eta \in \{0.05, 0.1, 0.15, 0.2, 0.3\}$. Protocol I recovers the exact graph in 9 of 10 runs for every $\eta$ between 0.15 and 0.3, in 7 of 10 at $\eta = 0.1$ and in none at $\eta = 0.05$. Protocol O recovers it for no threshold; its smallest mean SHD is 2.2, at $\eta = 0.2$. The default $\eta = 0.1$ is thus conservative for I and does not cause the failure of O.

\paragraph{Correlation sweep}
The protocols O$_a$ of Section~\ref{sec:ident-results} were run with three seeds each. Attribution errors per run fall from 1.2 (O, $\rho = -0.998$) to 1.0 ($-0.967$), 0.33 ($-0.822$) and 0 ($-0.698$), as the certificate predicts. The two other error types persist, however, at 1.7 to 2.0 per run, and no O$_a$ run recovers the exact graph. Lower correlation alone therefore does not make an observational design as good as protocol I. In O$_a$, $\what$ remains an affine function of $\That$ within each specimen, and all tests share the same four temperature levels. In protocol I, the moisture gradient is orthogonal to the temperature gradient within each specimen. We read this as follows. Correlation controls the attribution errors, and the Fisher certificate captures that. Under noise, a design can still be biased, and designs that decorrelate the environment within each specimen were more robust here. The certificate is local and blind to bias.

\paragraph{$\sigma_u = 3\times 10^{-5}$}
Both protocols fail in all five runs (mean SHD 3.8 for O, 4.0 for I). The error is now dominated by bias of type 2. Noise in the strains biases the viscous branch strongly: $E_1^0$ drops to about 0.7 (truth 1.5) and $\tau^0$ rises to 35--39\,s (truth 20\,s). Spurious environmental edges on $E_1$ and $\kappa_0$ then compensate the bias. The RMS strain noise is about $2.5\times 10^{-4}$ at $\sigma_u = 10^{-5}$ and three times that here, comparable to the eigenstrains of about $10^{-3}$ that carry edges 8 and 9. Interventions cannot correct an errors-in-variables bias; Section~\ref{sec:discussion} discusses remedies.

\subsection{Interventional prediction}\label{sec:prediction}

\begin{figure}[t]
\centering
\includegraphics[width=\textwidth]{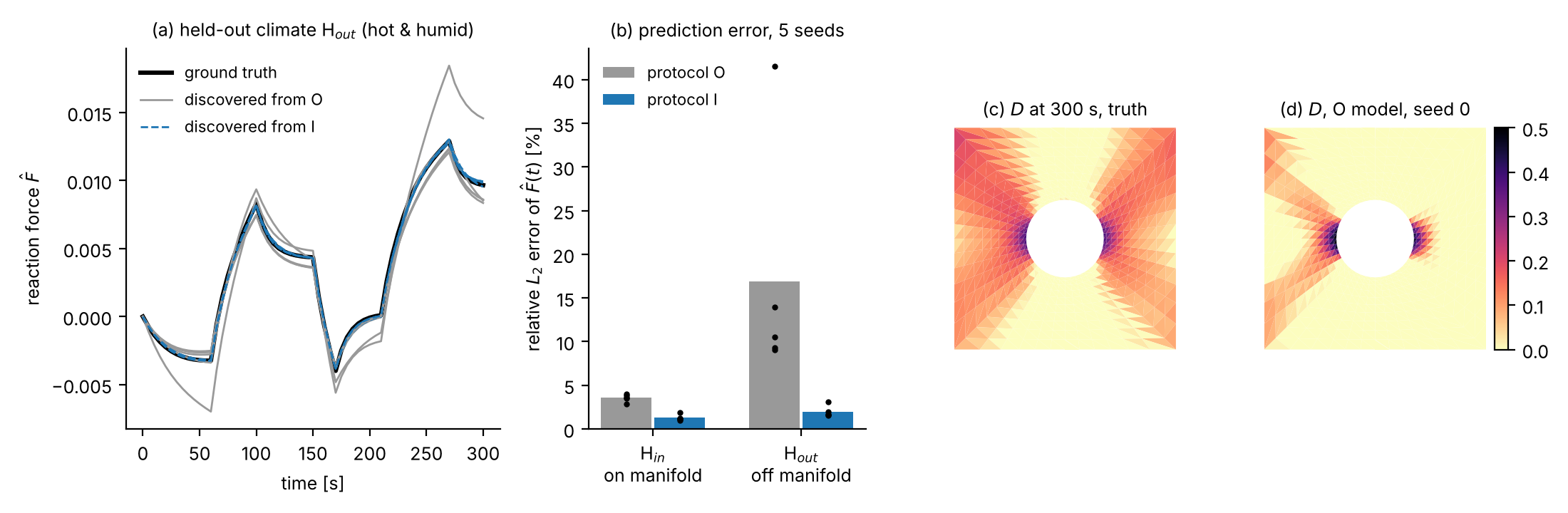}
\caption{Interventional prediction with models discovered at $\sigma_u = 10^{-5}$. (a) Reaction-force history for the held-out hot and humid climate H$_{\mathrm{out}}$. (b) Relative $L_2$ error of the reaction-force history; bars show the mean, dots the five seeds. (c, d) Damage at $t = 300$\,s for H$_{\mathrm{out}}$, ground truth and the worst observational model.}
\label{fig:prediction}
\end{figure}

Models discovered from observational data predict a climate on their own manifold with 3.6\,\% mean error. A hot and humid climate they mispredict by 16.9\,\% on average and by up to 41.5\,\%. Models discovered from interventional data stay within 3.2\,\% in both cases (Fig.~\ref{fig:prediction}). We forward-simulated the held-out tests H$_{\mathrm{in}}$ and H$_{\mathrm{out}}$ with the Newton solver, using the ten sparse models of seeds 0--4 at $\sigma_u = 10^{-5}$, and compared the reaction-force histories with the ground truth.

\paragraph{On the manifold (H$_{\mathrm{in}}$, hot and dry)}
The O-models reach relative $L_2$ errors of 2.8--4.0\,\% (mean 3.6\,\%), the I-models 1.0--1.9\,\% (mean 1.3\,\%). An engineer who validates the O-models on reaction forces under a training-like climate would accept them.

\paragraph{Off the manifold (H$_{\mathrm{out}}$, hot and humid)}
The errors of the O-models rise to 9.0--41.5\,\% (mean 16.9\,\%), while those of the I-models remain at 1.6--3.1\,\% (mean 2.0\,\%). Fig.~\ref{fig:prediction}a shows where the O-models fail. Already in the conditioning phase they predict the wrong eigenstress, because they split the combined expansion between $T$ and $w$ incorrectly. They also misjudge peak and relaxed forces, because the moisture effects on $\Einf$ and $\kappa_0$ are attributed to temperature. The worst O-model (seed 0) underestimates the mean final damage by a factor of 2.4 (0.036 against 0.084) and predicts a markedly different damage pattern (Fig.~\ref{fig:prediction}c,d). Another (seed 2) overestimates it (0.150).

\paragraph{Latent fields}
Reaction forces hide errors in the latent fields, also for the I-models. On H$_{\mathrm{in}}$, all O-models overestimate the mean final damage by 49--66\,\% (0.084--0.094 against 0.056). On H$_{\mathrm{out}}$, four I-models underestimate it by 23--29\,\% (0.060--0.065 against 0.084), and one overestimates it by 13\,\%. The cause is the bias of the damage parameters discussed in Section~\ref{sec:recovery}: $D_{\max} = 0.51$--$0.68$ and $\kappa_f = 0.009$--$0.017$ under I, against 0.5 and 0.01. A correct graph does not guarantee correct latent fields at this noise level.

\paragraph{Full-graph baseline}
To test whether sparsity causes the failure of O, we repeated the predictions with the unpenalised stage-1 models, in which all 13 edges are active. The O-models then reach 2.9--3.6\,\% on H$_{\mathrm{in}}$ and 12.5--18.4\,\% (mean 13.9\,\%) on H$_{\mathrm{out}}$. The I-models reach 2.0--2.3\,\% on both. The failure of O is therefore a property of the data, not of the selection. For I, sparsity slightly improves the predictions (mean 1.3\,\% against 2.2\,\% on H$_{\mathrm{in}}$).

This experiment is the mechanical counterpart of a familiar warning from causal inference: predictive error on the training distribution says little about the response to an intervention~\cite{Pearl2009,Schoelkopf2021}. Here the intervention is a new climate, and its consequences are not small.

\subsection{Acyclicity and lagged feedback}\label{sec:ablation}

\begin{figure}[t]
\centering
\includegraphics[width=0.62\textwidth]{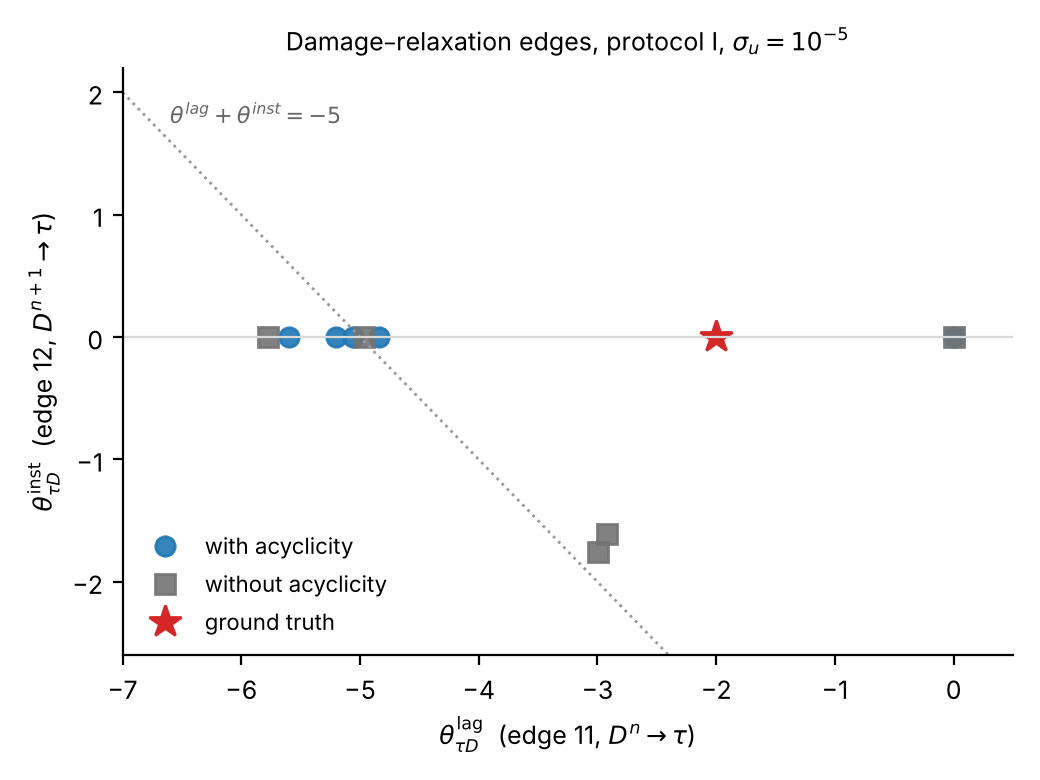}
\caption{Lagged and instantaneous damage--relaxation edges discovered from protocol I at $\sigma_u = 10^{-5}$, with and without the acyclicity constraint (seeds 0--4 each, equal iteration budget). Seed 3 lies at the origin in both settings. The dotted line marks a constant total coupling $\theta^{\lag} + \theta^{\inst} = -5$.}
\label{fig:ablation}
\end{figure}

Without the acyclicity constraint, two of five runs keep a spurious instantaneous loop $\bvarepsilon_v \to D \to \tau \to \bvarepsilon_v$; with it, none do. We repeated seeds 0--4 of protocol I at $\sigma_u = 10^{-5}$ with $\mu = \rho = 0$ and without the 2-cycle projection (Fig.~\ref{fig:ablation}). The iteration budget was the same ($3\times 80$ L-BFGS iterations in stage 2), and the local loop was converged with $N_{\mathrm{fp}} = 8$ sweeps.

Without the constraint, two runs split the damage effect on relaxation between the two edges: $\theta^{\inst}_{\tau D} = -1.61$ and $-1.75$, with a residual acyclicity $h$ of 0.25 and 0.31. Their sums $\theta^{\lag} + \theta^{\inst}$ are $-4.5$ and $-4.8$. In the remaining runs the instantaneous weight vanished without the constraint. With the constraint, the instantaneous weight is exactly zero in all runs, and the lagged weight lies between $-4.8$ and $-5.6$ in four of them. The data thus fix the total damage--relaxation coupling but not its timing. This is the near-null direction of Section~\ref{sec:ident-results}.

The constraint does not improve the fit. Among equally good models it selects the one whose material update has an explicit order of evaluation (Remark~\ref{rem:order}). Exact recoveries were 4 of 5 with the constraint and 2 of 5 without it; the difference consists only of the instantaneous edge. Seed 3 behaves identically in both settings: it loses the lagged edge and keeps the spurious $T \to E_1$ (SHD 2). A first version of this ablation used a smaller iteration budget and two unconverged sweeps without the constraint. It retained the loop in four of five runs, which overstated the effect.

The magnitude of the coupling is biased by noise: around $-5$ instead of the true $-2.0$, whereas noise-free data return $-2.02$ (Section~\ref{sec:recovery}).

\subsection{Intervention design}\label{sec:design}

\begin{figure}[t]
\centering
\includegraphics[width=\textwidth]{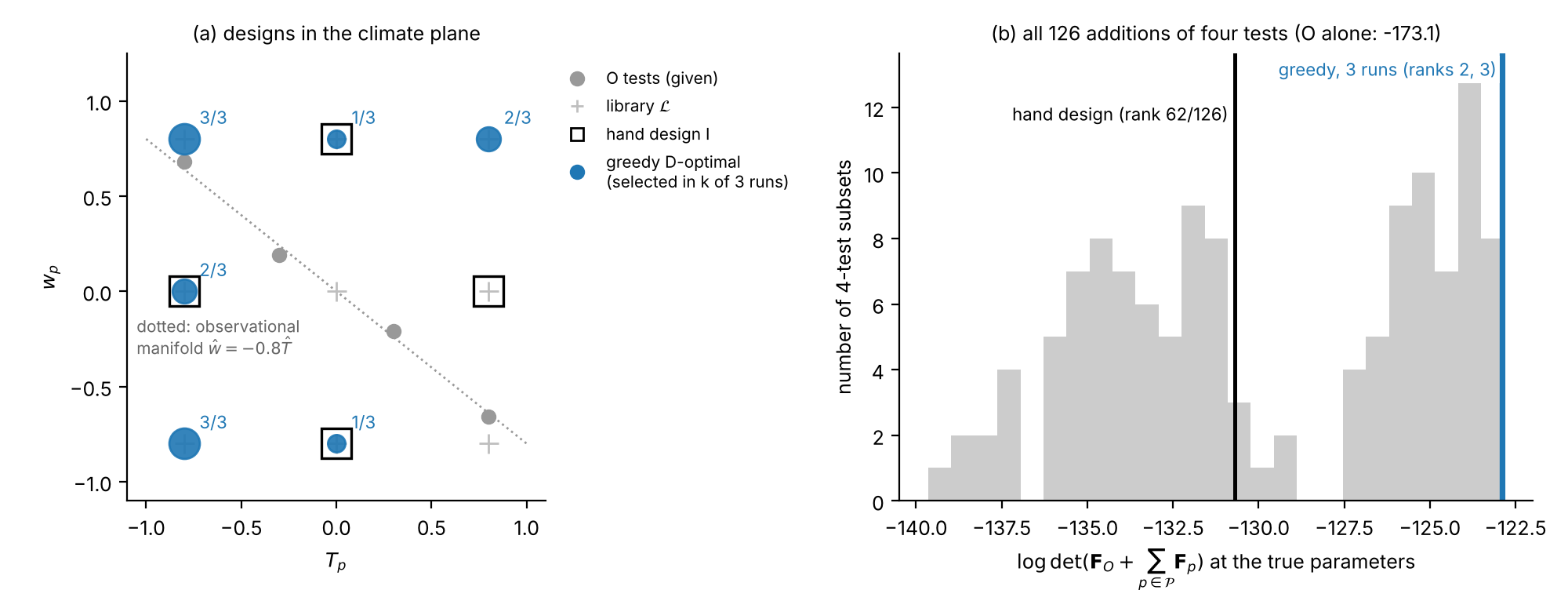}
\caption{Intervention design. (a) Observational tests, library, hand design and greedy D-optimal selections in the plane of target temperature and moisture; the marker size and label show in how many of three design runs (different starting models) a test was selected. (b) $\log\det$ of the Gauss--Newton matrix at the true parameters, common reference force, for all 126 ways of adding four library tests to protocol O.}
\label{fig:design}
\end{figure}

The greedy D-optimal design, started from the observational tests alone, ranks second or third among all 126 possible four-test additions, for three different starting models. The hand-designed protocol I ranks 62nd (Fig.~\ref{fig:design}).

We took the four tests of protocol O as given and used a model discovered from them at $\sigma_u = 10^{-5}$ as the current estimate $\hat\btheta$. Each of the nine library tests was forward-simulated with $\hat\btheta$, and its Gauss--Newton matrix was evaluated at $\hat\btheta$ from these predicted fields; the ground truth was not used. The predicted fields stand in for data that do not yet exist. This is what makes the design sequential, and it is accurate to the extent that $\hat\btheta$ predicts the new tests. We repeated the design with the models of seeds 0, 1 and 2. The greedy selections were $\{(-0.8,-0.8), (0.8,0.8), (-0.8,0.8), (-0.8,0)\}$ twice and $\{(-0.8,-0.8), (-0.8,0.8), (0,-0.8), (0,0.8)\}$ once, written as $(T_p, w_p)$. All contain the corner $(-0.8,-0.8)$, which lies farthest from the observational manifold $\what = -0.8\,\That$ together with $(0.8, 0.8)$, as Proposition~\ref{prop:collinear} suggests.

To judge the designs after the fact, we evaluated $\log\det(\bF_O + \sum_p \bF_p)$ at the true parameters for all 126 subsets, with the common reference force of Section~\ref{sec:identifiability}. The value is $-173.1$ for O alone. Adding the hand design gives $-130.7$ (rank 62, at the median of $-131.0$). The greedy designs give $-122.92$ and $-122.85$ (ranks 3 and 2), within 0.09 of the best subset ($-122.83$). The difference of 7.8 between greedy and hand design corresponds to a volume ratio of the local confidence ellipsoids of $\exp(7.8/2) \approx 50$. The hand design is a one-factor-at-a-time star; the library itself is the $3\times 3$ factorial.

A first version of this study used a per-test normalisation of the residuals instead of the common reference force. It ranked the greedy design first and the hand design 71st, while the protocol-wide normalisation of the score ranked them 75th and 28th. The criterion is therefore only meaningful with a normalisation that does not change with the subset, as used here.

The gain in information did not translate into fewer structural errors at $\sigma_u = 10^{-5}$. We repeated the discovery with the four O tests plus one of three designs, three seeds each. The designs were the greedy design of the first version, the hand design and a design of intermediate rank, ranked 7, 62 and 20 under the common normalisation. All nine runs recovered every true edge, and eight of them included the spurious edge $T \to E_1$ (SHD 1); the ninth, with the hand design, reached SHD 0. Compared with O alone (mean SHD 3.2), adding four off-manifold tests removes all attribution errors, although this comparison also doubles the data. The remaining error is the noise-induced bias of type 2 ($E_1^0 \approx 1.26$--$1.35$ against 1.5). D-optimality cannot address it, because it targets variance, not bias. The design gain may become visible in edge recovery once the bias is removed, or at the margin of identifiability, for instance with fewer or shorter tests. Three seeds per design are too few to detect smaller differences.

\subsection{Computational cost}\label{sec:cost}

One evaluation of the score for four tests takes 16\,ms; score and gradient together take 0.14\,s on a two-core CPU (JAX, double precision). The Gauss--Newton matrix of one test costs 0.08\,s with forward-mode differentiation over the 19 parameters. A complete three-stage discovery took 150--400\,s of wall time, with two or three runs sharing the two cores; noisier data needed more line-search steps. For comparison, one Newton forward simulation of a single test (60 steps, up to 15 iterations with line search, dense tangent) takes 29\,s. Avoiding forward solves inside the optimisation is therefore essential: the three stages evaluate the score and its gradient several hundred times.

\section{Discussion and limitations}\label{sec:discussion}

The experiments support one central message: in constitutive model discovery, the test protocol limits what can be learned more than the algorithm does. A sparse, well-fitting model is no evidence of correct mechanism attribution. Interventions, and a computable certificate of what they make identifiable, belong in the discovery pipeline. The study also shows the limits of that certificate, and several further limitations qualify the message.

\paragraph{Variance versus bias}
The Gauss--Newton certificate predicts the attribution errors caused by collinear parents, and these disappear when the environment is decorrelated. It does not predict the bias that measurement noise induces in the weak-form score. That bias produced a spurious viscous edge and the loss of a weak damage edge in every observational-type design, even at $|\rho| = 0.70$, but rarely under protocol I. A certificate that also accounts for bias is needed. Options are the Fisher information of the measurement model, obtained by implicit differentiation of the forward solver, or a comparison of discovered models across designs.

\paragraph{Measurement noise}
The weak-form score uses strains computed from noisy displacements. Noise therefore enters the regressors, not only the residual, and biases the estimates, most visibly in the viscous branch ($E_1$, $\tau$) and in the damage parameters. At $\sigma_u = 3\times 10^{-5}$ this bias produced false edges under every protocol. We used independent nodal noise. Real DIC noise is spatially correlated over a subset, which lowers the strain noise for a given displacement noise but introduces correlated errors; this needs to be tested. Remedies include denoising of the displacement fields as in EUCLID~\cite{Flaschel2021}, total-least-squares or output-error formulations such as finite element model updating, which avoid strains as regressors~\cite{Avril2008}, and a Bayesian treatment~\cite{Joshi2022}.

\paragraph{Synthetic ground truth and model class}
All data were generated by the same model class that is searched, and the truth is log-linear and additive in $T$ and $w$. Real materials show interactions, for example time--temperature--moisture superposition. An additive model identified from a one-factor-at-a-time design such as protocol I cannot represent interactions, and the held-out climate H$_{\mathrm{out}}$ lies outside the convex hull of protocol I as well. The excellent extrapolation of the I-models thus relies on additivity. Interaction candidates ($\That\what$ terms) and deliberately misspecified truths are the next tests. With real data, a missing mechanism will be absorbed by the candidate edges that correlate best with it. An explicit discrepancy term, or a test for residual structure, is needed before discovered edges can be read causally.

\paragraph{Candidate set and causal sufficiency}
Discovery is confined to the 13 candidate edges. The method can reject an edge, but it cannot propose a mechanism outside the library. Richer hypothesis spaces, such as grammar-generated expressions or neural sub-mechanisms with sparse inputs, fit into the same framework, because the score is differentiable in any parameterisation. Unmeasured variables that co-vary with the environment, such as ageing, curing or conditioning history, act as hidden confounders. Interventions on $T$ and $w$ do not remove them. Randomising the order of tests and checking invariance across specimens~\cite{Peters2016} are natural safeguards. Such hidden confounders are also where the causal formulation goes beyond collinear regression.

\paragraph{What acyclicity does and does not decide}
The lagged and the instantaneous damage edge are indistinguishable at the time resolution used here (Section~\ref{sec:ident-results}). The acyclicity constraint selects the lagged edge by assumption, namely that the material update has an explicit order of evaluation (Remark~\ref{rem:order}). It is not an empirical finding. If the distinction matters physically, it must be resolved by experiments with time steps comparable to the damage--relaxation coupling.

\paragraph{Exogeneity of the environment}
We treated $T$ and $w$ as exogenous because the chamber sets them. In reality, dissipation heats the specimen and cracks accelerate moisture uptake, so $T$ and $w$ become partly endogenous. The time-unrolled graph accommodates such edges as long as they act with a lag, and the acyclicity constraint then becomes more important, not less.

\paragraph{Cost}
One discovery run with four tests took 150--400\,s on a two-core CPU (Section~\ref{sec:cost}). The score requires no linear solve, so its cost grows linearly with mesh size and number of steps. A forward solver is needed only for prediction and for designing the next experiment.

\paragraph{Relevance for structural engineering}
Design codes describe environmental effects on materials through modification factors that act on fixed mechanisms. Examples are service and load-duration classes in timber design, or temperature- and duration-dependent interlayer stiffness in laminated glass. Which mechanism such a factor should modify is exactly the parent-set question studied here. An interventional, certified discovery could support the calibration of such factors from laboratory campaigns, instead of fitting them to observational climate data.

\section{Conclusions}\label{sec:conclusions}

We have formulated constitutive model discovery as causal discovery of mechanism parent sets. Hard-coded physics and learnable mechanism edges form a time-unrolled structural causal model, and candidate graphs are scored through the boundary value problem by a differentiable weak-form equilibrium gap. The main findings of the synthetic study are:
\begin{enumerate}[leftmargin=*]
\item At a DIC-typical noise of about 0.01\,px, climate-chamber interventions recover the true mechanism graph exactly in 7 of 10 runs and make no $T$/$w$ attribution error in any run. Observational data with a temperature--moisture correlation of $-0.998$ never recover it (mean SHD 3.2) and misattribute the moisture effect on the damage threshold in every run.
\item The attribution errors are predicted by a computable certificate. Along the collinear directions of Proposition~\ref{prop:collinear}, the score of the observational protocol is 250--380 times flatter, of the order of the variance inflation factor. When the correlation is lowered, attribution errors vanish as the certificate predicts.
\item The certificate is necessary but not sufficient. Noise biases the weak-form score, and this bias caused a spurious viscous edge and the loss of a weak damage edge in all observational-type designs, but rarely under the climate-chamber design.
\item Fit is no proof of mechanism. Observationally discovered models predict a climate on their training manifold within 3.6\,\% on average, but a hot and humid climate with 16.9\,\% mean and 41.5\,\% maximum error. An unpenalised full-graph fit fails similarly, so the failure lies in the data, not in the selection. Interventionally discovered models stay within 3.2\,\%. They still misestimate latent damage by up to about 30\,\%.
\item Acyclicity of the instantaneous graph makes the material update explicit in its order of evaluation, and it removed a spurious algebraic loop that two of five unconstrained runs retained.
\item Greedy D-optimal design from observational data alone selected test sets ranked second or third of 126 under a consistent normalisation, for three different starting models. An intuitive one-factor-at-a-time design ranked 62nd.
\end{enumerate}
Three limitations stand out. Noise of 0.03\,px biases the viscous branch enough to create false edges under any protocol. The truth is additive in $T$ and $w$, so the extrapolation results do not cover interactions. And the candidate library bounds what can be discovered. The next steps are bias-aware scores, either output-error formulations or the measurement-model Fisher information by implicit differentiation, together with interaction and grammar-based hypothesis spaces and validation on climate-chamber experiments with DIC on polymers and laminated-glass interlayers.

\section*{Code and data availability}
The JAX implementation (mesh, structural causal model, Newton solver, weak-form score, discovery algorithm, Fisher analysis and experiment scripts) and all synthetic data will be released in a public repository upon publication.

\section*{CRediT authorship contribution statement}
\textbf{Michael Kraus:} Conceptualization, Methodology, Software, Investigation, Writing -- original draft, Writing -- review \& editing.

\section*{Declaration of generative AI in the writing process}
During the preparation of this work the author used AI, specifically Anthropic Claude Opus 5.5, openAI gpt-6-astra in its ChatGPT version as well as Google gemini to draft code, text and figures. The author reviewed, iterated and edited all content and takes full responsibility for the publication.

\section*{Declaration of competing interest}
The author declares no competing financial interests or personal relationships that could have appeared to influence the work reported in this paper.

\bibliographystyle{elsarticle-num}
\bibliography{refs}

\end{document}